\documentclass[aps,prl,twocolumn,amsmath,showpacs,superscriptaddress]{revtex4}

\usepackage{amsmath}
\usepackage{amsbsy}
\usepackage{amscd}
\usepackage{amsopn}
\usepackage{amstext}
\usepackage{amsxtra}
\usepackage{epsfig}

\def\kp{{k^\prime}}

\def\lesssim{\mathrel{\hbox{\rlap{\hbox{\lower4pt\hbox{$\sim$}}}\hbox{$<$}}}}

\begin{document}

\title{Bond percolation on a class of clustered random networks}
\author{James P. Gleeson}

\affiliation{Department of Mathematics \& Statistics, University
of Limerick, Ireland.}

\date{5  Aug 2009: ver6}

\begin{abstract}
Analytical results are derived for the bond percolation threshold
 and the size of the giant connected component in a class of
random networks with non-zero clustering. The network's degree
distribution and clustering spectrum may be prescribed, and
theoretical results match well to numerical simulations on both
synthetic and real-world networks.
\end{abstract}

\maketitle

Random network models have been extensively studied with a view to
gaining insight into the structure and dynamics of many social,
technological, and biological
networks~\cite{Newman03a,Dorogovtsev03,Dorogovtsev08}. However,
most analytical approaches rely on tree-like approximations of the
local network structure and thus neglect the presence of short
loops (cycles) in the graphs. The \emph{local clustering
coefficient} for a node $A$ is defined as the fraction of pairs of
neighbors of node $A$ which are also neighbors of each
other~\cite{Watts98}, and is typically non-negligible in
real-world networks. The \emph{degree-dependent clustering} or
\emph{clustering spectrum} $c_k$ is the average of the local
clustering coefficient over the class of all nodes of degree
$k$~\cite{Serrano06a,Vazquez02}. The question of how network
models with  non-zero $c_k$ (taken, for example, from real-world
network data) differ from randomly-wired
(\emph{configuration-model}) networks with the same degree
distribution $P_k$  is of considerable interest.

The bond percolation problem for a network may be stated as
follows: each edge of the network graph is visited once, and
\emph{damaged} (deleted) with probability $1-p$. The quantity $p$
is the \emph{bond occupation probability} and the non-damaged
edges are termed \emph{occupied}. The size of the \emph{giant
connected component} (GCC) of the graph  becomes nonzero at some
critical value of $p>0$: this critical value of $p$ is termed the
\emph{bond percolation threshold} $p_{th}$. The bond percolation
problem has applications in epidemiology, where $p$ is related to
the average transmissibility of a disease and the GCC represents
the size of an epidemic outbreak \cite{Grassberger83,Newman02b},
and in the analysis of technological networks, where the
resilience of a network to the random failure of links is
quantified by the size of the GCC~\cite{Serrano06b}. Analytical
solutions for percolation on randomly-wired
%(\emph{configuration
%model})
networks and on correlated networks are well-known
\cite{Molloy95,Callaway00,Newman01a,Vazquez03}, but these cases
have zero clustering in the limit of infinite network size.

In this paper we introduce a class of networks with non-zero
clustering, and demonstrate analytical solutions for the GCC size
and the bond percolation threshold. Most previous studies of
clustering effects on percolation rely on numerical simulations
using various algorithms to generate clustered networks, e.g.
\cite{Klemm02,Volz04,Serrano05}. Analytical solutions were found
by Newman~\cite{Newman03b} for a bipartite graph model of highly
clustered networks. However, the bipartite graph model (in
contrast to the model discussed here) is not amenable to fitting
to a prescribed degree distribution $P_k$. The bipartite graph
model of Guillaume and Latapy \cite{Guillaume06} may be fitted to
real-world data but their networks do not permit analytical
solution of the percolation problem. Serrano and
Bogu\~{n}\'{a}~\cite{Serrano06c, Serrano06b} also obtain
approximate analytical solutions, but only for weak clustering
cases with $c_k<1/(k-1)$. Trapman~\cite{Trapman07} introduced a
model of clustering in structured graphs based on embedding
cliques (complete subgraphs) within a random tree structure. We
show below that this model, and its generalization
\cite{Gleeson08c} are in fact special cases of the model presented
here. In a recent paper \cite{Newman09}, Newman introduced a
triangle-based model of clustered networks which may be seen as
complementary to the model presented here: we discuss this model
in detail at the end of the paper.
\begin{figure}
\centering
\includegraphics[width=0.75\columnwidth]{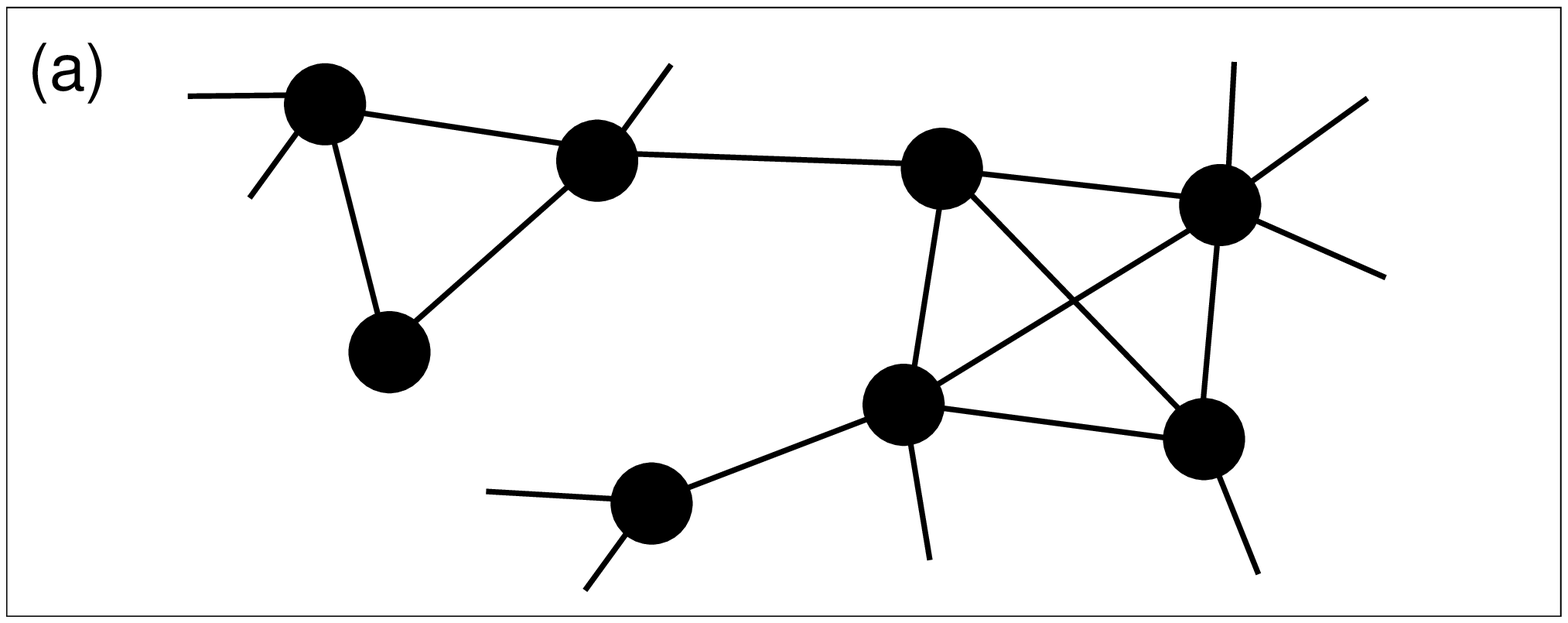}\\
\includegraphics[width=0.75\columnwidth]{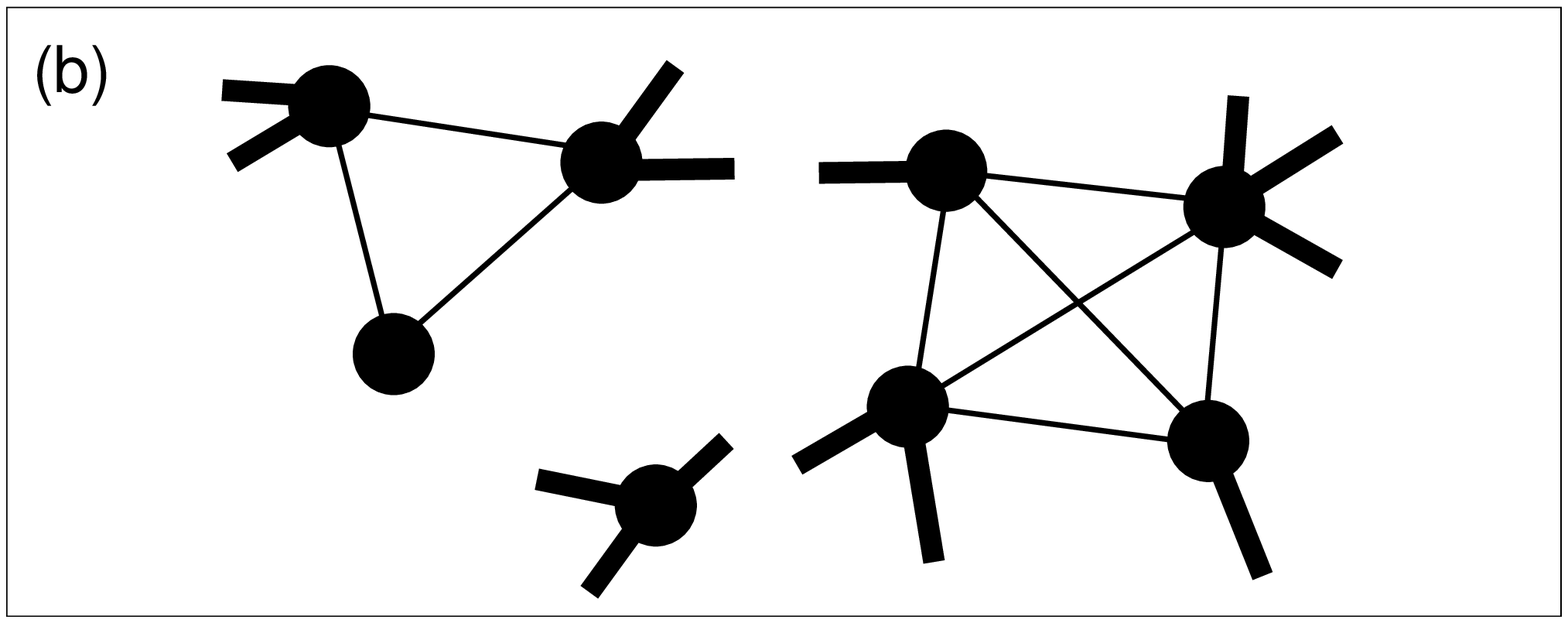}\\
\includegraphics[width=0.75\columnwidth]{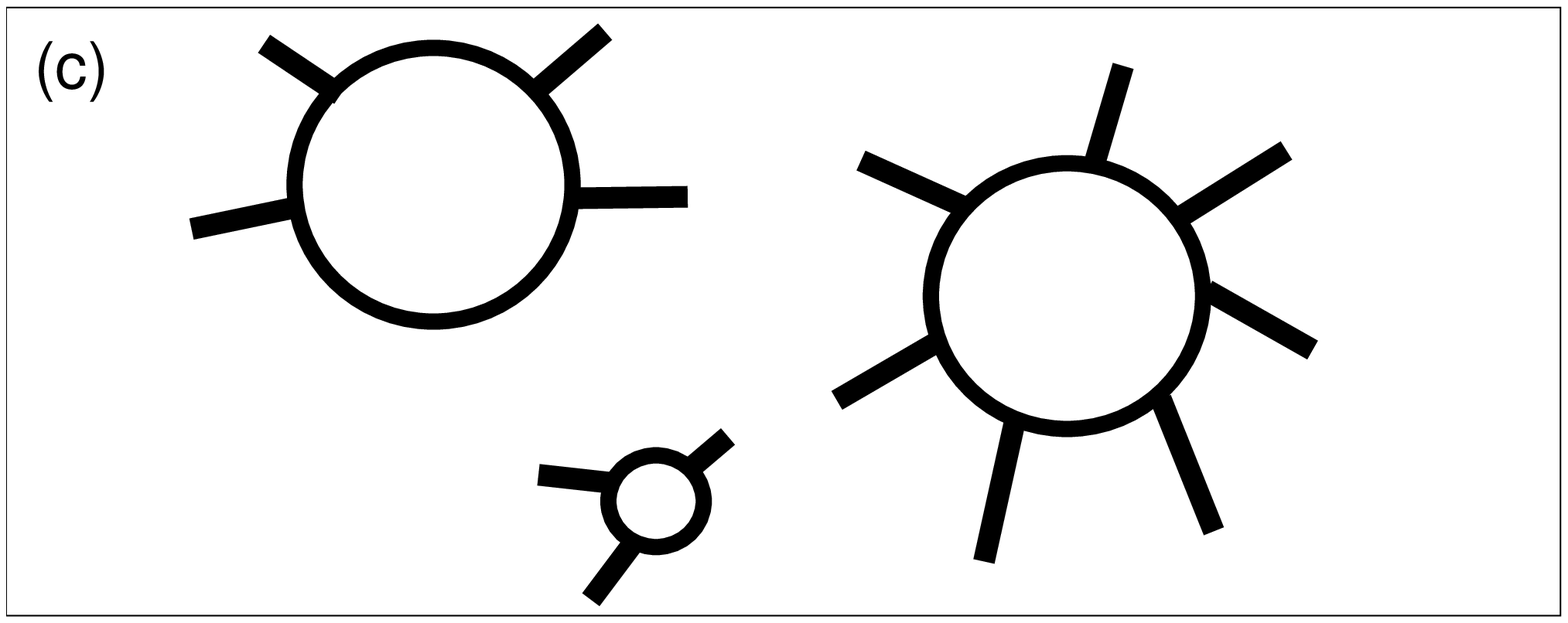}\\
%\vspace{0.0cm}
 \caption{ (a) Segment of a clustered random network;
 (b) split into disjoint cliques, with external links emphasized; (c) corresponding super-nodes.}
 \label{fig1}
\end{figure}

We consider  random networks in which each node  may be part of a
single clique (a fully-connected subgraph).
 Figure~\ref{fig1}(a)
shows a segment of such a network which contains one 3-clique
(triangle), one 4-clique, and a single node which is not a member
of a clique (for notational convenience we will refer to such
individual nodes as members of a 1-clique). Nodes which are
members of a $c$-clique have $c-1$ edges linking them to neighbors
within the same clique. They also have  an additional $k-c+1$
neighbors who are not in the same clique as themselves,  where $k$
is the node degree (with
 $k\ge c-1$).  Edges
which are not internal to a clique are termed \emph{external
links}. In Fig.\ref{fig1}(b) the external links are highlighted
with thick lines, but for the purposes of the bond percolation
problem they are indistinguishable from clique edges. In networks
of this type each node is a member of at most one clique, and so
the network can be decomposed into disjoint cliques which are
linked together by the set of external links, see
Fig.~\ref{fig1}(b) \footnote{We differentiate between
\emph{external links} and 2-cliques as follows. An external link
joins together two nodes, each of which may be part of its own
clique, e.g., at the top of Fig.\ref{fig1}(a) an external link
joins a 3-clique node to a 4-clique node. A 2-clique is also an
edge joining two nodes, but because these nodes are in the
2-clique they cannot be part of any other clique and so can link
to the remainder of the network only through external links.}. If
each clique is regarded as a \emph{super-node}
(Fig.~\ref{fig1}(c)) then realizations of the random network may
be generated by connecting together randomly chosen pairs of the
external link stubs, as in the configuration model for standard
random networks \cite{Newman01a}.
%An algorithm for generating
%realizations of networks of this type is described in detail
%below.

The fundamental quantity describing networks of this type is the
joint probability distribution $\gamma(k,c)$, giving the
probability that a randomly-chosen node in the network has degree
$k$ and is a member of a $c$-clique.
%The normalization is
%$\sum_{k=0}^\infty \sum_{c=1}^{k+1} \gamma(k,c) = 1$, which can be
%abbreviated to $\sum_{k,c} \gamma(k,c) = 1$ since the probability
%is zero for nodes with
Note $\gamma(k,c)=0$ for $k<c-1$, i.e., $k$-degree nodes can only
be members of $c$-cliques if their degree is high enough to
provide links to all $c-1$ clique neighbors. The degree
distribution $P_k$ of the network (probability that a random node
has $k$ neighbors) is obtained from $\gamma$ by averaging over all
cliques:
\begin{equation}
P_k = \sum_{c=1}^{k+1} \gamma(k,c) = \sum_c \gamma(k,c).
\label{Peqn}
\end{equation}
A node chosen at random from the set of all $k$-degree nodes is a
member of a $c$-clique with probability  $\gamma(k,c)/P_k$. As a
member of a $c$-clique, it is part of $\left( \!\!\begin{array}{c}
c-1
\\ 2 \end{array}\!\! \right)$ triangles, and so its local
clustering coefficient is $\left( \!\!\begin{array}{c} c-1
\\ 2 \end{array}\!\! \right)/\left( \!\begin{array}{c}
k
\\ 2 \end{array}\! \right)$. Therefore the degree-dependent
clustering coefficient $c_k$ is given in terms of $\gamma$ by
\begin{equation}
c_k = \sum_c \frac{\gamma(k,c)}{P_k} \frac{(c-1)(c-2)}{k(k-1)}.
\label{Ceqn}
\end{equation}

The class of networks  described by the joint pdf $\gamma(k,c)$
includes the well-studied configuration model \cite{Newman01a},
for which $\gamma(k,c) = \delta_{c 1} P_k$. This limit contains no
cliques, and hence the clustering (in the infinite network size
limit) vanishes. Also contained within the class of networks is
the Trapman model \cite{Trapman07,Gleeson08c} in which a fraction
$f_k$ of $k$-degree nodes form cliques of precisely $k$ nodes,
giving $\gamma(k,c) = \delta_{c 1}(1-f_k) P_k + \delta_{c k} f_k
P_k$.
%The case where $f_k=F$ for all $k$ was studied in
%\cite{Trapman07} and generalized to arbitrary $f_k$ in
%\cite{Gleeson08c}.

\begin{figure}
\centering
\includegraphics[width=0.75\columnwidth]{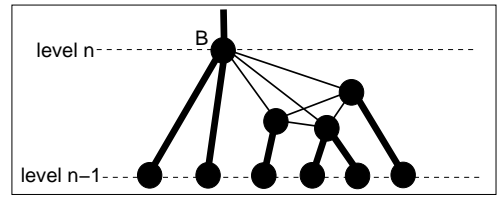}
 \caption{ Tree diagram for updating the state of node $B$.}
 \label{fig2}
\end{figure}
To determine the expected size of the GCC in the damaged network
we choose a random node $A$ of the network and approximate the
network as a tree structure, with the node $A$ at the top (root)
of the tree. Each level of the tree structure (see
Fig.~\ref{fig2}) is accessed from the level above by traversing
one external link. If a node is part of a $c$-clique, the
remaining $c-1$ clique neighbors are shown at an intermediate
level.
%Figure~\ref{fig2} shows, for example, a node of degree 6
%which is in a 4-clique, with one external link to the level above.
%It also has 2 external links to the level below, and is connected
%to 4 further external links to the level below via its clique
%neighbors.
Because the graph of super-nodes (Fig.~\ref{fig1}(c)) is connected
using the configuration model, this tree structure is a locally
accurate approximation to the original network is the limit of
infinite system size.
%%(provided that the degree distribution of the super-nodes has finite variance).
%(short loops appear in the graph of super-nodes with a
%probability that vanishes as $N\to\infty$).

To calculate the probability that node $A$ is part of the GCC, we
apply a tree-based approach  which is generalizable to a variety
of cascade dynamics on networks \cite{Gleeson08a} and is related
to work on the random field Ising model \cite{Dhar97}. We label
nodes which are part of a connected component as \emph{active}
with the remaining nodes termed \emph{inactive}. All nodes of the
tree are initially considered inactive, and we examine the
propagation of the active state upwards through the tree (from
leaves to root) as an infection process beginning from an
infinitesimally small fraction of active nodes infinitely deep in
the tree.
%We
%track the fraction of active nodes as we move from level to level
%up the infinite tree towards the root node $A$.
Consider a node at
level $n$, e.g. node $B$ in Fig.~\ref{fig2}. Initially $B$ (and
its parent at level $n+1$) is inactive, but suppose nodes at level
$n-1$ are active with probability $q$.
%The probability of $B$
%becoming active is determined from the dynamics of the bond
%percolation process.
If $B$ has degree $k$, and is a member of a $c$-clique (e.g. $k=6$
and $c=4$ in Fig.~\ref{fig2}), it has $k-c+1$ external links, one
of which necessarily leads to its parent at level $n+1$.
%One external link
%leads to $B$'s parent at level $n+1$, and there are $k-c$ other
%external links which directly connect (if occupied) to level-$n$
%children of $B$.
The node $B$ will become active if any one of its $k-c$
externally-linked  children at level $n-1$ is active, provided
that an occupied edge joins that child  to $B$; thus the
probability that $B$ is \emph{not} activated in this fashion is
%$1-(1-p q)^{k-c}$.
$(1-p q)^{k-c}$. The other mechanism whereby $B$ may be activated
is via its neighbors in the $c$-clique; writing $Q_c$ for the
probability that the top-node (such as $B$) of a $c$-clique is
activated by its clique neighbors, we have the total probability
of activation for $B$ of $
1-(1-p q)^{k-c}(1-Q_c) %\label{A1}.
$. The probability $Q_c$ is calculated using the $P(m|k)$
functions introduced and tabulated in \cite{Newman03b}, which are
polynomials in $p$ giving the
% where
%an iterative procedure gives $P(m|c)$, the
probability that a randomly chosen node in a damaged (i.e., taking
into account bond percolation) $c$-clique belongs to a connected
cluster of $m$ nodes (including itself) within the clique. Since
node $B$ is activated if any one of its $m-1$ connected neighbors
is active, we have
\begin{equation}
Q_c = \sum_{m=1}^c P(m|c)\left( 1-
(1-\overline{q}_c)^{m-1}\right), \label{A2} \end{equation} where
$\overline{q}_c$ is the probability of a $c$-clique member at the
intermediate level being activated by his level-$(n-1)$ children:
\begin{equation}
\overline{q}_c = \sum_{\kp}\frac{\gamma(\kp,c)
}{\sum_{k^{\prime\prime}}\gamma(k^{\prime\prime},c)} \left(1-(1- p
q )^{k^\prime-c+1} \right) \label{A3}.
\end{equation}
Here ${\gamma(\kp,c) }/
{\sum_{k^{\prime\prime}}\gamma(k^{\prime\prime},c)}$ is the degree
distribution of nodes which are members of cliques of size $c$,
and the remaining term is the probability that a $\kp$-degree node
in a $c$-clique is activated by one of its $\kp-c+1$ children at
level $n-1$.

Given $q$, we can therefore calculate, using equations (\ref{A2})
and (\ref{A3}), the probability of $B$ becoming active. To close
the system of equations, we consider the parent of $B$ at level
$n+1$, for whom $q$ is  the probability that one of its children
is active.
%(in the steady-state of this infection-type process).
%,
%and on an infinite network).
Since node $B$ has $k-c+1$ external links in total, the
probability of it being a child of a random level-$(n+1)$ node is
$\Pi_{k c}= (k-c+1)\gamma(k,c)/z_e$, where
$z_e=\sum_{\kp,c^\prime}(\kp-c^\prime+1)\gamma(\kp,c^\prime)$ is
the average number of external links per node. Combining the
equations above gives the closure relation
\begin{equation}
q=\sum_{k,c} \Pi_{k c}\left(1-(1-p q)^{k-c}(1-Q_c)\right) \equiv
G(q) \label{A4}.
\end{equation}
Equations (\ref{A2})--(\ref{A4}) are solved by iterating from an
infinitesimally small value of $q$ to a steady-state solution:
this determines the probability (in an infinite network) that a
node is active, conditional on its parent being inactive. The
final calculation of the GCC size considers the node $A$ at the
top (root) of the tree: with probability $\gamma(k,c)$ this has
 $k-c+1$ direct links to
the level below. By similar arguments to before, node $A$ is
active with probability
\begin{equation}
S = \sum_{k,c} \gamma(k,c) \left(1 - (1-p q
)^{k-c+1}(1-Q_c)\right), \label{A5}
\end{equation}
where $q$ is the solution of equations (\ref{A2})--(\ref{A4}) and
$S$ is the expected fractional size of the GCC.

Note that if we set $\gamma(k,c) = \delta_{c 1} P_k$, equations
(\ref{A4})  and (\ref{A5}) reduce to their well-studied
configuration model versions (using $Q_1=0$).
% This reduction (via
%the notation mapping $1-p q \mapsto x$) recovers, for example,
%equations (9) and (14) of \cite{Dorogovtsev08}.
The bond percolation results of \cite{Trapman07,Gleeson08c} for
the generalized Trapman model are also a special case of equations
(\ref{A2})--(\ref{A5}). Also of interest is the GCC size in an
undamaged network---this is obtained from our equations by setting
$p=1$ (for which $P(m|c)=\delta_{m c}$).
% and hence reducing
%(\ref{A2}) to $Q_c = 1-(1-\overline{q}_c)^{c-1}$.

The bond percolation threshold is the value of $p$ at which the
GCC size first becomes nonzero. This may be determined from the
cascade condition \cite{Gleeson08c,Gleeson08a} $G^\prime(0)=1$,
where $G(q)$ is defined in equation (\ref{A4}). The resulting
polynomial equation for $p$ may be written in the form
\begin{equation}
 \frac{1}{z_e}\sum_{k,c} (k-c+1) \gamma(k,c)\left(p(k-c) +
(z_c-c+1)D_c(p)\right) = 1 \label{gperc},
\end{equation}
where $D_c(p) = p \sum_{m=1}^c (m-1) P(m|c)$ (see
\cite{Gleeson08c}) and $z_c$ is the average degree of nodes in
cliques of size $c$: $z_c = \sum_k k \gamma(k,c) /\sum_{\kp}
\gamma(\kp,c)$.

%Detailed study of the corresponding equation for the special case
%of the generalized Trapman model  has shown that the presence of
%clustering can cause either an increase or a decrease of the bond
%percolation threshold relative to its value in an unclustered
%network with the same degree distribution $P_k$ and may even lead
%to a non-zero percolation threshold in scale-free networks for
%which $P_k$ has infinite variance \cite{Trapman07,Gleeson08c}.

We now describe an algorithm for generating realizations of random
networks with a prescribed distribution $\gamma(k,c)$. For a large
number $\widetilde{N}$ (which is related to the number $N$ of
nodes in the final network, see below), we choose $\widetilde N$
random numbers $c_i$ ($i=1$ to $\widetilde N$) with pdf
$\left(\sum_k \gamma(k,c)/c\right)/\left(\sum_{k,c^\prime}
\gamma(k,c^\prime)/c^\prime\right)$ to be the clique sizes in the
network realization.
%Each $c_i$ value corresponds to the size of a
%$c$-clique in the final network, and hence to $c_i$ nodes.
For each $c_i$, we create $c_i$ nodes in a complete subgraph, and
assign their degrees $k_j$ ($j=1$ to $c_i$) by drawing random
numbers from a distribution with density
$\gamma(k,c_i)/\sum_{\kp}\gamma(\kp,c_i)$. Node $j$ in clique $i$
then has $k_j-c_i+1$ external link stubs associated with it.
Having created all $\widetilde N$ cliques in this fashion, we
randomly choose pairs of external link stubs and connect them
together to create the random network (c.f. Fig.~\ref{fig1}). The
expected number of nodes in a network generated using this
algorithm is $N=\widetilde
N/\left(\sum_{k,c}\gamma(k,c)/c\right)$, which allows us to
estimate the value of $\widetilde N$ needed to produce a final
network of size $N$. For finite-sized networks, the presence of
cliques means this algorithm is not guaranteed to give exactly $N$
nodes in the final network, but in practice we find the variation
in the network size is negligibly small for sufficiently large
$N$.

\begin{figure}
\centering
\includegraphics[width=0.75\columnwidth]{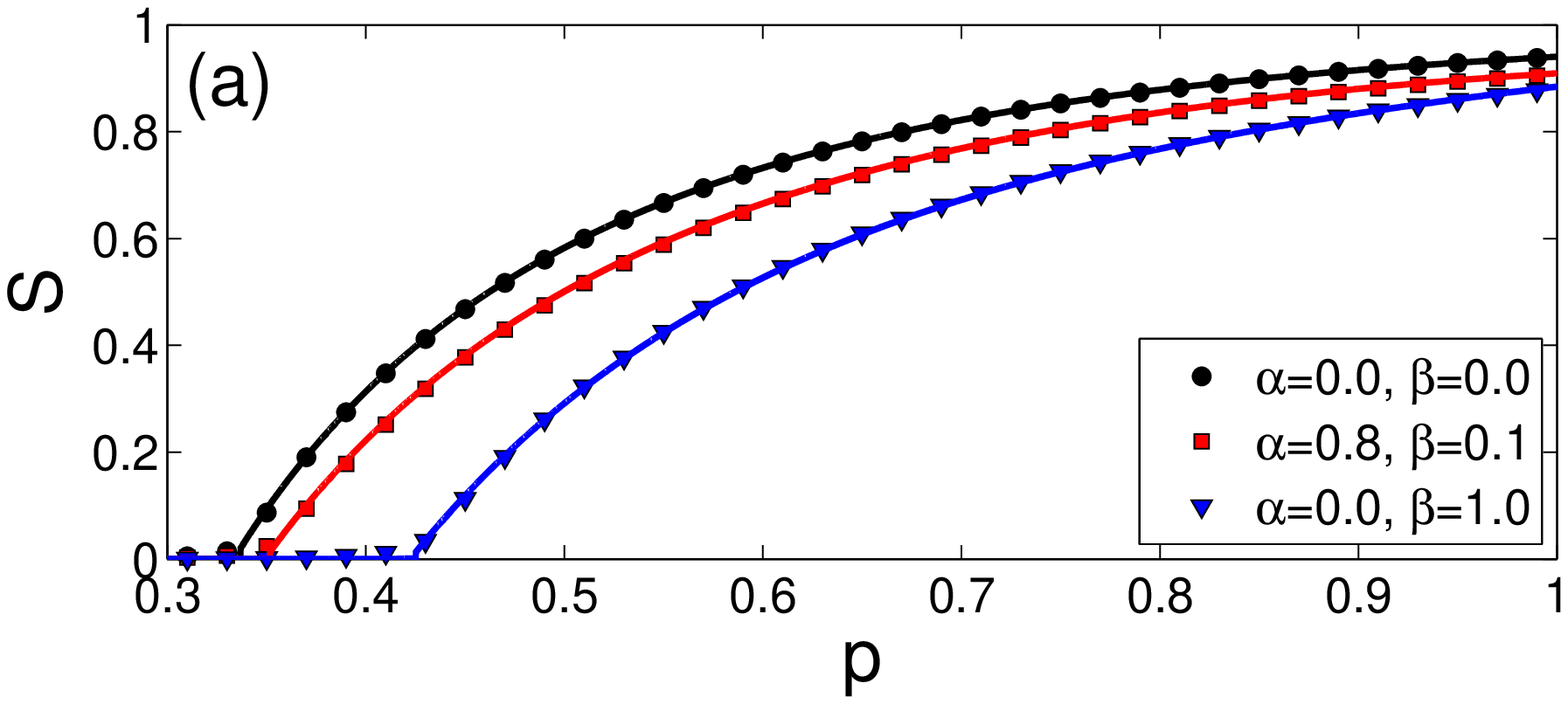}\\
\includegraphics[width=0.75\columnwidth]{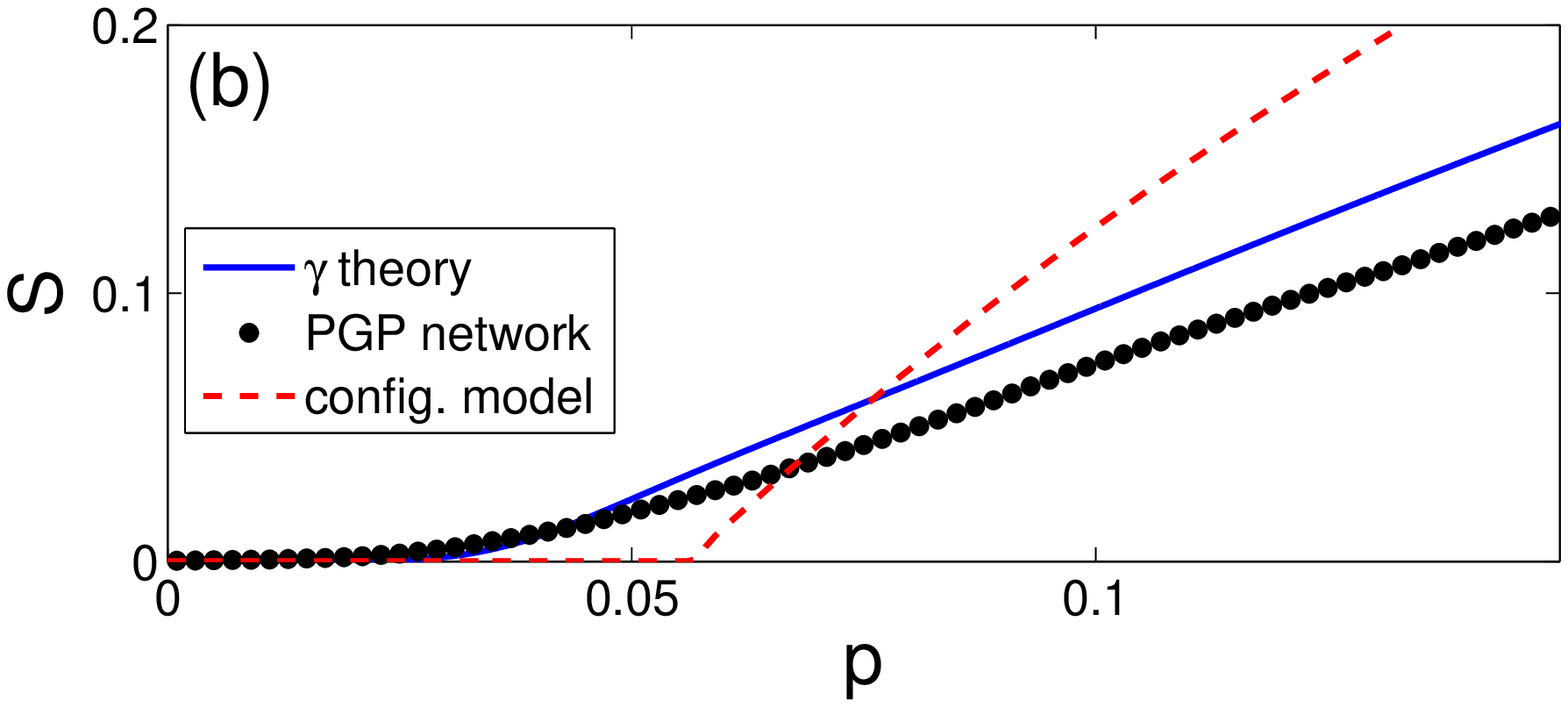}\\
\includegraphics[width=0.75\columnwidth]{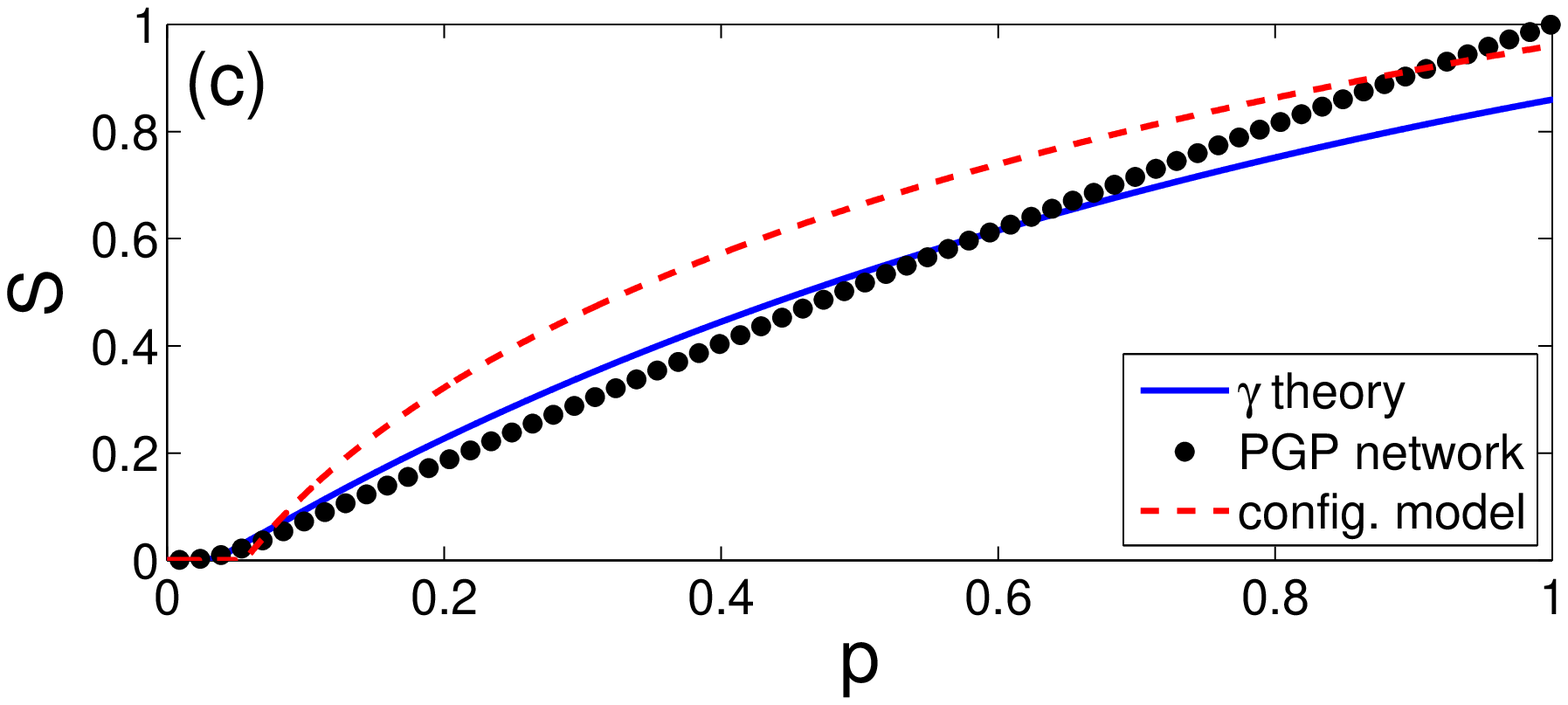}
%\vspace{0.2cm}
 \caption{ (Color online) Size of giant connected component as a function
 of bond occupation probability for (a) synthetic networks with Poisson degree distribution;
 (b) pretty-good-privacy network, for $p$ near $p_{th}$; (c) as (b), but for all $p$ values.}
  \label{fig3}
\end{figure}
Figure~\ref{fig3}(a) shows a comparison between GCC sizes from
theory (from equations (\ref{A2})--(\ref{A5})) and numerical
simulations on networks with the Poisson degree distribution $P_k
= z^k e^{-z}/k!$ and mean degree $z=3$.
%Realizations of networks
%with a prescribed $\gamma(k,c)$ distribution are generated by
%first drawing a set of random cliques-sizes from the appropriate
%marginal distribution, and then allocating the degrees of the
%nodes in all cliques; this algorithm will be described in detail
%in a subsequent paper.
%For Fig.~\ref{fig3}(a) w
We create non-zero clustering in the networks by inserting
3-cliques (triangles) and 4-cliques; specifically, we set
$\gamma(k,c) = \left( (1-\alpha-\beta) \delta_{c 1} + \alpha
\delta_{c 3} + \beta \delta_{c 4}\right) P_k$ for $k \ge 3$. This
embeds a fraction $\alpha$ and $\beta$ of $k$-degree nodes in
3-cliques and 4-cliques, respectively, with the remainder as
individuals (i.e, 1-cliques). Since nodes of degree $k$ cannot be
part of $c$-cliques when $c$ exceeds $k+1$, we deal with nodes of
degree $k<3$ as follows: $\gamma(2,c) = \left((1-\alpha)\delta_{c
1} + \alpha \delta_{c 3}\right) P_2$, and $\gamma(k,c) = P_k
\delta_{c 1}$ for $k=0$ or 1. The case $\alpha=\beta=0$ gives the
standard configuration model network, with zero clustering. We
also show results for $\alpha=0.8$, $\beta=0.1$ and for
$\alpha=0$, $\beta=1$. Using (\ref{Ceqn}), the first of these
corresponds to a global clustering coefficient $C=\sum_k P_k c_k$
of 0.31, while the second case, which contains only 4-cliques, has
$C=0.35$. The corresponding bond percolation thresholds may be
calculated from the polynomial equation (\ref{gperc}) using (see
\cite{Gleeson08c}) $D_3(p)= 2 p^2(1+p-p^2)$ and $D_4(p) = 3 p^2
(1+2 p -7p^3+7p^4-2 p^5)$. The resulting values are $p_{th}=0.349$
and $p_{th}=0.422$, both exceeding the configuration model value
of $p_{th} = 1/z = 1/3$ \cite{Molloy95,Callaway00}. Numerical
simulation results on networks of size $N=10^5$
%(averaged over 10 realizations)
are shown by the
symbols, while the curves are the theoretical predictions of
equations (\ref{A2})--(\ref{A5}). The agreement between theory and
numerics is excellent.

One of the motivations for the introduction of the $\gamma$
networks is the ability to obtain analytical results for networks
with given degree distribution $P_k$ and clustering spectrum
$c_k$. Equations (\ref{Peqn}) and (\ref{Ceqn}) constrain the
distribution $\gamma(k,c)$ to fit a desired $P_k$ and $c_k$, which
may be measured, for example, in a real-world network. However,
these constraints still permit significant freedom in choosing
$\gamma(k,c)$. It is convenient therefore to consider a
parametrization of $\gamma(k,c)$ which allows straightforward
fitting to given network data. We suppose that the distribution of
clique-sizes $c$ occupied by nodes of degree $k$ is given by a
binomial distribution, defining
\begin{equation}
\gamma(k,c) = P_k \,\left(\!\! \begin{array}{c} k\\c-1
\end{array}\!\! \right) g_k^{c-1}(1-g_k)^{k-c+1} \label{gparam},
\end{equation}
for $c=1$ to $k+1$. This distribution clearly satisfies
(\ref{Peqn}), and it distributes the probability mass
corresponding to the $k$-degree nodes over the $c$-clique sizes
via the single parameter $g_k$.
%The mean size of cliques occupied
%by $k$-degree nodes is $\sum_c c \gamma(k,c)/P_k = k g_k +1 $,
%with $g_k=0$ giving the unclustered case $\gamma(k,c) = \delta_{c
%1 } P_k$.
The relationship between the parameters $g_k$ and the clustering
spectrum $c_k$ is remarkably simple; substituting the
parametrization (\ref{gparam}) into (\ref{Ceqn}) yields $c_k =
g_k^2$. Thus the form (\ref{gparam}) for $\gamma$ may immediately
be fitted to the $P_k$ and $c_k$ of a real-world dataset by
setting $g_k = \sqrt{c_k}$.

%\begin{figure}
%\centering
%\includegraphics[width=0.9\columnwidth]{fig_draw_fig_theory.eps}
% \caption{ (Color online)CAption caption caption caoption caption caption caption } \label{fig4}
%\end{figure}
Figures~\ref{fig3}(b) and \ref{fig3}(c) show the results of
applying the parametrization (\ref{gparam}) to match the degree
distribution  and clustering spectrum  of the connected component
of the pretty-good-privacy (PGP) network \cite{Boguna04}. The PGP
network is highly clustered, with $c_k>1/(k-1)$ for most $k$
\cite{Serrano06b}.
% This network
%contains ?? nodes, with mean degree $z=$ , maximum degree
%$k_text{max}= $, and has $c_k$ values of approximately   ?? with
%global clustering coefficient $C= \sum_k P_k c_k = $.
Numerical calculations of the GCC size in this network are shown
by the symbols on Fig.~\ref{fig3}(b) and \ref{fig3}(c); also shown
are the theoretical predictions for the zero-clustering
(configuration model) case and the results of equations
(\ref{A2})--(\ref{A5}) with parametrization (\ref{gparam}). The
effects  of clustering on the percolation threshold are
well-captured by the $\gamma$ theory, see Fig.~\ref{fig3}(b). Note
that here, in contrast to the example in Fig.~\ref{fig3}(a),
clustering acts to decrease the percolation threshold: equation
(\ref{gperc}) gives $p_{th}=0.0236$, which is less than half the
configuration model value of $0.0559$. The $\gamma$ theory gives
quite a good approximation to the actual GCC size for bond
occupation probabilities of $p$ up to about 0.5
(Fig.~\ref{fig3}(c)); however the behavior at larger $p$ values is
less accurate, with the predicted GCC in the undamaged ($p=1$)
network being substantially smaller than its true value. This
inaccuracy may be attributable to excessive clustering being
induced by the parametrization (\ref{gparam}) for the large $p$
case.
%, and alternative parametrizations are under consideration.
%, and alternative parameterizations
%are also under investigation.

In summary, we have introduced a class of clustered random
networks with arbitrary degree distribution and clustering
spectrum, and analytically determined the size of the GCC and the
bond percolation threshold. Numerically generated networks
%generated by our algorithm
show excellent agreement with the theoretical results, and we have
demonstrated the applicability of the theory by fitting to the
pretty-good-privacy network to produce accurate predictions of the
GCC size for small $p$. We have used a cascade-based approach here
in preference to a generating function method, because (as we show
in a subsequent paper) this approach generalizes to give
analytical results for $k$-core sizes, Watts' threshold decision
model, and other cascading dynamics on clustered networks
\cite{Gleeson08a}.

It is instructive to compare our $\gamma$-theory networks with the
clustered network model recently introduced by Newman
\cite{Newman09}. In his model, a $k$-degree node may be a member
of up to $k/2$ disjoint triangles (3-cliques), and thus have a
local clustering coefficient of up to $1/(k-1)$. In contrast,
nodes in the $\gamma$-theory networks can be members of only a
single clique, but using large cliques can give arbitrarily high
clustering.
%  but
%since the clique may be large the local clustering coefficient may
%be arbitrarily close to 1.
The restriction $c_k \le 1/(k-1)$ imposed on Newman's model
networks inhibits a direct fit to most real-world networks, in
contrast to our results in Fig.~\ref{fig3}. It would be
interesting to explore the possibility of modelling networks with
multiple cliques per node (as in \cite{Newman09}) while allowing
the cliques to be larger than triangles (as here). Indeed, a
general model of this type is proposed in \cite{Bollobas08} but it
seems unlikely that easily computable analytical solutions, as
found here and in \cite{Newman09}, can be obtained in this more
general setting.

Discussions with Sergey Melnik, Adam Hackett and Mason Porter are
gratefully acknowledged. This work was funded by Science
Foundation Ireland under programmes 06/IN.1/I366 and MACSI
06/MI/005.

\bibliography{networks}

\begin{thebibliography}{26}
\expandafter\ifx\csname natexlab\endcsname\relax\def\natexlab#1{#1}\fi
\expandafter\ifx\csname bibnamefont\endcsname\relax
  \def\bibnamefont#1{#1}\fi
\expandafter\ifx\csname bibfnamefont\endcsname\relax
  \def\bibfnamefont#1{#1}\fi
\expandafter\ifx\csname citenamefont\endcsname\relax
  \def\citenamefont#1{#1}\fi
\expandafter\ifx\csname url\endcsname\relax
  \def\url#1{\texttt{#1}}\fi
\expandafter\ifx\csname urlprefix\endcsname\relax\def\urlprefix{URL }\fi
\providecommand{\bibinfo}[2]{#2}
\providecommand{\eprint}[2][]{\url{#2}}

\bibitem[{\citenamefont{Newman}(2003{\natexlab{a}})}]{Newman03a}
\bibinfo{author}{\bibfnamefont{M.~E.~J.} \bibnamefont{Newman}},
  \bibinfo{journal}{SIAM Rev.} \textbf{\bibinfo{volume}{45}},
  \bibinfo{pages}{167} (\bibinfo{year}{2003}{\natexlab{a}}).

\bibitem[{\citenamefont{Dorogovtsev and Mendes}(2003)}]{Dorogovtsev03}
\bibinfo{author}{\bibfnamefont{S.}~\bibnamefont{Dorogovtsev}} \bibnamefont{and}
  \bibinfo{author}{\bibfnamefont{J.}~\bibnamefont{Mendes}},
  \emph{\bibinfo{title}{Evolution of Networks: From Biological Nets to the
  Internet and WWW}} (\bibinfo{publisher}{Oxford University Press, Oxford},
  \bibinfo{year}{2003}).

\bibitem[{\citenamefont{Dorogovtsev et~al.}(2008)\citenamefont{Dorogovtsev,
  Goltsev, and Mendes}}]{Dorogovtsev08}
\bibinfo{author}{\bibfnamefont{S.~N.} \bibnamefont{Dorogovtsev}},
  \bibinfo{author}{\bibfnamefont{A.~V.} \bibnamefont{Goltsev}},
  \bibnamefont{and} \bibinfo{author}{\bibfnamefont{J.~F.~F.}
  \bibnamefont{Mendes}}, \bibinfo{journal}{Rev. Mod. Phys.}
  \textbf{\bibinfo{volume}{80}}, \bibinfo{pages}{1275} (\bibinfo{year}{2008}).

\bibitem[{\citenamefont{Watts and Strogatz}(1998)}]{Watts98}
\bibinfo{author}{\bibfnamefont{D.~J.} \bibnamefont{Watts}} \bibnamefont{and}
  \bibinfo{author}{\bibfnamefont{S.~H.} \bibnamefont{Strogatz}},
  \bibinfo{journal}{Nature (London)} \textbf{\bibinfo{volume}{393}},
  \bibinfo{pages}{440} (\bibinfo{year}{1998}).

\bibitem[{\citenamefont{Serrano and
  Bogu\~{n}\'{a}}(2006{\natexlab{a}})}]{Serrano06a}
\bibinfo{author}{\bibfnamefont{M.~{\relax \'A}.} \bibnamefont{Serrano}}
  \bibnamefont{and}
  \bibinfo{author}{\bibfnamefont{M.}~\bibnamefont{Bogu\~{n}\'{a}}},
  \bibinfo{journal}{Phys. Rev. E} \textbf{\bibinfo{volume}{74}},
  \bibinfo{pages}{056114} (\bibinfo{year}{2006}{\natexlab{a}}).

\bibitem[{\citenamefont{V\'{a}zquez et~al.}(2002)\citenamefont{V\'{a}zquez,
  Pastor-Satorras, and Vespignani}}]{Vazquez02}
\bibinfo{author}{\bibfnamefont{A.}~\bibnamefont{V\'{a}zquez}},
  \bibinfo{author}{\bibfnamefont{R.}~\bibnamefont{Pastor-Satorras}},
  \bibnamefont{and}
  \bibinfo{author}{\bibfnamefont{A.}~\bibnamefont{Vespignani}},
  \bibinfo{journal}{Phys. Rev. E} \textbf{\bibinfo{volume}{65}},
  \bibinfo{pages}{066130} (\bibinfo{year}{2002}).

\bibitem[{\citenamefont{Grassberger}(1983)}]{Grassberger83}
\bibinfo{author}{\bibfnamefont{P.}~\bibnamefont{Grassberger}},
  \bibinfo{journal}{Math. Biosci.} \textbf{\bibinfo{volume}{63}},
  \bibinfo{pages}{157} (\bibinfo{year}{1983}).

\bibitem[{\citenamefont{Newman}(2002)}]{Newman02b}
\bibinfo{author}{\bibfnamefont{M.~E.~J.} \bibnamefont{Newman}},
  \bibinfo{journal}{Phys. Rev. E} \textbf{\bibinfo{volume}{66}},
  \bibinfo{pages}{016128} (\bibinfo{year}{2002}).

\bibitem[{\citenamefont{Serrano and
  Bogu\~{n}\'{a}}(2006{\natexlab{b}})}]{Serrano06b}
\bibinfo{author}{\bibfnamefont{M.~{\relax \'A}.} \bibnamefont{Serrano}}
  \bibnamefont{and}
  \bibinfo{author}{\bibfnamefont{M.}~\bibnamefont{Bogu\~{n}\'{a}}},
  \bibinfo{journal}{Phys. Rev. E} \textbf{\bibinfo{volume}{74}},
  \bibinfo{pages}{056115} (\bibinfo{year}{2006}{\natexlab{b}}).

\bibitem[{\citenamefont{Molloy and Reed}(1995)}]{Molloy95}
\bibinfo{author}{\bibfnamefont{M.}~\bibnamefont{Molloy}} \bibnamefont{and}
  \bibinfo{author}{\bibfnamefont{B.}~\bibnamefont{Reed}},
  \bibinfo{journal}{Random Structures and Algorithms}
  \textbf{\bibinfo{volume}{6}}, \bibinfo{pages}{161} (\bibinfo{year}{1995}).

\bibitem[{\citenamefont{Callaway et~al.}(2000)\citenamefont{Callaway, Newman,
  Strogatz, and Watts}}]{Callaway00}
\bibinfo{author}{\bibfnamefont{D.~S.} \bibnamefont{Callaway}},
  \bibinfo{author}{\bibfnamefont{M.~E.~J.} \bibnamefont{Newman}},
  \bibinfo{author}{\bibfnamefont{S.~H.} \bibnamefont{Strogatz}},
  \bibnamefont{and} \bibinfo{author}{\bibfnamefont{D.~J.} \bibnamefont{Watts}},
  \bibinfo{journal}{Phys. Rev. Lett.} \textbf{\bibinfo{volume}{85}},
  \bibinfo{pages}{5468} (\bibinfo{year}{2000}).

\bibitem[{\citenamefont{Newman et~al.}(2001)\citenamefont{Newman, Strogatz, and
  Watts}}]{Newman01a}
\bibinfo{author}{\bibfnamefont{M.~E.~J.} \bibnamefont{Newman}},
  \bibinfo{author}{\bibfnamefont{S.~H.} \bibnamefont{Strogatz}},
  \bibnamefont{and} \bibinfo{author}{\bibfnamefont{D.~J.} \bibnamefont{Watts}},
  \bibinfo{journal}{Phys. Rev. E} \textbf{\bibinfo{volume}{64}},
  \bibinfo{pages}{026118} (\bibinfo{year}{2001}).

\bibitem[{\citenamefont{V\'{a}zquez and Moreno}(2003)}]{Vazquez03}
\bibinfo{author}{\bibfnamefont{A.}~\bibnamefont{V\'{a}zquez}} \bibnamefont{and}
  \bibinfo{author}{\bibfnamefont{Y.}~\bibnamefont{Moreno}},
  \bibinfo{journal}{Phys. Rev. E} \textbf{\bibinfo{volume}{67}},
  \bibinfo{pages}{015101(R)} (\bibinfo{year}{2003}).

\bibitem[{\citenamefont{Klemm and Eguiluz}(2002)}]{Klemm02}
\bibinfo{author}{\bibfnamefont{K.}~\bibnamefont{Klemm}} \bibnamefont{and}
  \bibinfo{author}{\bibfnamefont{V.~M.} \bibnamefont{Eguiluz}},
  \bibinfo{journal}{Phys. Rev. E} \textbf{\bibinfo{volume}{65}},
  \bibinfo{pages}{036123} (\bibinfo{year}{2002}).

\bibitem[{\citenamefont{Volz}(2004)}]{Volz04}
\bibinfo{author}{\bibfnamefont{E.}~\bibnamefont{Volz}}, \bibinfo{journal}{Phys.
  Rev. E} \textbf{\bibinfo{volume}{70}}, \bibinfo{pages}{056115}
  (\bibinfo{year}{2004}).

\bibitem[{\citenamefont{Serrano and Bogu\~{n}\'{a}}(2005)}]{Serrano05}
\bibinfo{author}{\bibfnamefont{M.~{\relax \'A}.} \bibnamefont{Serrano}}
  \bibnamefont{and}
  \bibinfo{author}{\bibfnamefont{M.}~\bibnamefont{Bogu\~{n}\'{a}}},
  \bibinfo{journal}{Phys. Rev. E} \textbf{\bibinfo{volume}{72}},
  \bibinfo{pages}{036133} (\bibinfo{year}{2005}).

\bibitem[{\citenamefont{Newman}(2003{\natexlab{b}})}]{Newman03b}
\bibinfo{author}{\bibfnamefont{M.~E.~J.} \bibnamefont{Newman}},
  \bibinfo{journal}{Phys. Rev. E} \textbf{\bibinfo{volume}{68}},
  \bibinfo{pages}{026121} (\bibinfo{year}{2003}{\natexlab{b}}).

\bibitem[{\citenamefont{Guillaume and Latapy}(2006)}]{Guillaume06}
\bibinfo{author}{\bibfnamefont{J.-L.} \bibnamefont{Guillaume}}
  \bibnamefont{and} \bibinfo{author}{\bibfnamefont{M.}~\bibnamefont{Latapy}},
  \bibinfo{journal}{Physica A} \textbf{\bibinfo{volume}{371}},
  \bibinfo{pages}{795} (\bibinfo{year}{2006}).

\bibitem[{\citenamefont{Serrano and
  Bogu\~{n}\'{a}}(2006{\natexlab{c}})}]{Serrano06c}
\bibinfo{author}{\bibfnamefont{M.~{\relax \'A}.} \bibnamefont{Serrano}}
  \bibnamefont{and}
  \bibinfo{author}{\bibfnamefont{M.}~\bibnamefont{Bogu\~{n}\'{a}}},
  \bibinfo{journal}{Phys. Rev. Lett.} \textbf{\bibinfo{volume}{97}},
  \bibinfo{pages}{088701} (\bibinfo{year}{2006}{\natexlab{c}}).

\bibitem[{\citenamefont{Trapman}(2007)}]{Trapman07}
\bibinfo{author}{\bibfnamefont{P.}~\bibnamefont{Trapman}},
  \bibinfo{journal}{Theor. Pop. Biol.} \textbf{\bibinfo{volume}{71}},
  \bibinfo{pages}{160} (\bibinfo{year}{2007}).

\bibitem[{\citenamefont{Gleeson and Melnik}(2008)}]{Gleeson08c}
\bibinfo{author}{\bibfnamefont{J.~P.} \bibnamefont{Gleeson}} \bibnamefont{and}
  \bibinfo{author}{\bibfnamefont{S.}~\bibnamefont{Melnik}},
  \bibinfo{journal}{arXiv}  (\bibinfo{year}{2008}), \eprint{arXiv:0811.4511
  (submitted to Phys. Rev. E)}.

\bibitem[{\citenamefont{Newman}(2009)}]{Newman09}
\bibinfo{author}{\bibfnamefont{M.~E.~J.} \bibnamefont{Newman}},
  \bibinfo{journal}{Phys. Rev. Lett.} \textbf{\bibinfo{volume}{103}},
  \bibinfo{pages}{058701} (\bibinfo{year}{2009}).

\bibitem[{\citenamefont{Gleeson}(2008)}]{Gleeson08a}
\bibinfo{author}{\bibfnamefont{J.~P.} \bibnamefont{Gleeson}},
  \bibinfo{journal}{Phys. Rev. E} \textbf{\bibinfo{volume}{77}},
  \bibinfo{pages}{046117} (\bibinfo{year}{2008}).

\bibitem[{\citenamefont{Dhar et~al.}(1997)\citenamefont{Dhar, Shukla, and
  Sethna}}]{Dhar97}
\bibinfo{author}{\bibfnamefont{D.}~\bibnamefont{Dhar}},
  \bibinfo{author}{\bibfnamefont{P.}~\bibnamefont{Shukla}}, \bibnamefont{and}
  \bibinfo{author}{\bibfnamefont{J.~P.} \bibnamefont{Sethna}},
  \bibinfo{journal}{J. Phys. A} \textbf{\bibinfo{volume}{30}},
  \bibinfo{pages}{5259} (\bibinfo{year}{1997}).

\bibitem[{\citenamefont{Bogu\~{n}\'{a}
  et~al.}(2004)\citenamefont{Bogu\~{n}\'{a}, Pastor-Satorras, Diaz-Guilera, and
  Arenas}}]{Boguna04}
\bibinfo{author}{\bibfnamefont{M.}~\bibnamefont{Bogu\~{n}\'{a}}},
  \bibinfo{author}{\bibfnamefont{R.}~\bibnamefont{Pastor-Satorras}},
  \bibinfo{author}{\bibfnamefont{A.}~\bibnamefont{Diaz-Guilera}},
  \bibnamefont{and} \bibinfo{author}{\bibfnamefont{A.}~\bibnamefont{Arenas}},
  \bibinfo{journal}{Phys. Rev. E} \textbf{\bibinfo{volume}{70}},
  \bibinfo{pages}{056122} (\bibinfo{year}{2004}).

\bibitem[{\citenamefont{Bollobas et~al.}(2008)\citenamefont{Bollobas, Janson,
  and Riordan}}]{Bollobas08}
\bibinfo{author}{\bibfnamefont{B.}~\bibnamefont{Bollobas}},
  \bibinfo{author}{\bibfnamefont{S.}~\bibnamefont{Janson}}, \bibnamefont{and}
  \bibinfo{author}{\bibfnamefont{O.}~\bibnamefont{Riordan}},
  \bibinfo{journal}{arXiv}  (\bibinfo{year}{2008}), \eprint{arXiv:0807.2040}.

\end{thebibliography}
\end{document}